\documentclass{ws-ijmpd}

\begin{document}


\catchline{}{}{}{}{}

\title{Using supernova neutrinos to monitor the collapse, 
to search for gravity waves and to probe neutrino masses}

\author{F. VISSANI, G. PAGLIAROLI}\address{
INFN, Laboratori Nazionali del Gran Sasso, Theory Group, Assergi (AQ), Italy}

\author{F. ROSSI TORRES}\address{
Instituto de F\'isica ``Gleb Wataghin'' - UNICAMP, 13083-970 Campinas SP, Brazil\\INFN, Laboratori Nazionali del Gran Sasso, Theory Group, Assergi (AQ), Italy}

\maketitle

\begin{history}
\received{28 March 2010}
\end{history}

\begin{abstract}
We discuss the importance of observing supernova neutrinos. By analyzing the SN1987A observations of Kamiokande-II, IMB and Baksan, we show that they provide a $2.5\sigma$ support to the standard scenario for the explosion.  We discuss in this context the use of neutrinos as trigger for the search of the gravity wave impulsive emission. We derive a bound on the neutrino mass using the SN1987A data and argue, using simulated data, that a future galactic supernova could probe the sub-eV region.
\end{abstract}


\vskip5mm
\paragraph{Introduction}
Core collapse supernovae (type II, Ib and Ic) occur when the
progenitor has a mass $> 8\ M_\odot$ and originate compact
remnants: neutron stars, black holes and possibly hybrid
(=quark core) stars. The formation of such an object requires to carry away a
huge binding energy, several times $10^{53}$~erg. It is well known
that the role of carriers is played mainly by neutrinos; but more
importantly for us, and according to the standard scenario of core
collapse supernova explosion\cite{nad,Bethe:1984ux}, neutrinos
play also a fundamental role in driving the explosion. They
deposit energy that can revive the shock, which will eventually
cause the expulsion of the external layers of the star.

The current scenario of neutrino emission is based on two main
phases of neutrino emission.
The first one, called {\it accretion} phase, entails 10-20\%  of the total energy.
It is characterized by a very high neutrino luminosity and is directly related to
the matter which is accreted over the proto-neutron star through the stalled supernova shock wave.
The other phase is called {\it cooling} phase; the neutrinos escape slowly from the proto-neutron star,
releasing the remaining 80-90\% of the energy.
Only two analyses of SN1987A data included both emission phases: the analysis of 2001 by Loredo and
Lamb\cite{ll} and the recent one due to our group\cite{Pagliaroli:2008ur}. The most relevant
modification of this last analysis is the improvement of the model of $\bar \nu_e$
emission that we are going to describe in some detail in the
following.\footnote{Other features of the new
analysis: (i)~energy, time and direction of each event are taken
into account; (ii)~ the correct background\cite{jcap}, finite detection efficiency and
energy resolution are described; (iii)~dead times
and live time fraction are included; (iv) only the relative times
are used and the delay of the detector response, called {\it
offset times}, is accounted for; (v)~frequentist techniques of
inference are applied with an unbiased likelihood\cite{Ianni:2009bd}; and (vi)~the full $30$~s analysis window is
considered. An updated cross section of
IBD\cite{Strumia:2003zx} is used, an improved description of the
neutrino spectrum is introduced and neutrino
oscillations are accounted for in a suitable approximation.} Remarkably, both these
analyses claim an evidence of the phase of accretion in the
SN1987A data.

\paragraph{Detecting Supernova Neutrinos}
Before describing the model,  
we recall how many events
we expect from the most important detection reaction, the inverse beta
decay (IBD): $\bar \nu_e + p \rightarrow e^+ + n.$ 
 The number of events is: $N_{ev}=N_p \times
F_{\bar \nu_e} \times \sigma_{\bar\nu_e p}$.
The cross
section for a $\bar\nu_e$ with average energy $\bar{E}=15$~MeV is
$\sigma_{\bar\nu_e p} \approx G_F^2 \bar E^2 \approx
10^{-41}$~cm$^2$. The $\bar \nu_e$ fluence (time integrated flux)
can be written by
$
F_{\bar \nu_e} = \frac{{\cal E}_{b}/(6 \bar E)}{4\pi D^2} \sim \frac{2\cdot 10^{57}}{10(50\mbox{kpc})^2} \sim 10^{10} \frac{\bar \nu_e}{\mbox{cm}^2},
$
where ${\cal E}_{b}=3\cdot 10^{53}$~erg is the gravitational
binding energy, $D$ is the distance of the supernova and ``6''
are the neutrino types. For $1$~kton
detector, there are $ N_p \approx 1\mbox{ kton} \times 10^9 \times
6\cdot 10^{23} \times {2}/{18} \times 10^{32} $ protons. So the number of
expected events 
is about $10$. This rough
estimation agrees with the number of events observed:
$16$ in Kamiokande-II\cite{k} ($2140$~tons), $8$ in
IMB\cite{i} ($6800$~tons) and $5$ in Baksan\cite{b} ($200$~tons): $29$ events in 30
seconds, that include a few background events.

\begin{figure}[t]
\centerline{\includegraphics[width=.35\textwidth]{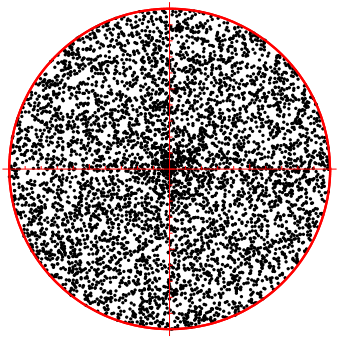}}
\caption{Directions of events events in a Cherenkov detector of 32 kton: the $\sim 300$ elastic scattering events are visible over the $\sim 5000$ inverse beta decay events, expected from the 
galactic center.
[We use the Lambert projection: the points of the unit sphere--i.e., the possible directions--identified by 
$n=(s_\theta c_\phi,s_\theta s_\phi, c_\theta)$, are mapped  
into the circle of radius 2, whose points 
are identified by $(u,v)=(c_\phi,s_\phi) \sqrt{2 (1-c_\theta)}$; namely, 
$u=n_x \sqrt{2/(1-n_z)}$ and $v=n_y \sqrt{2/(1-n_z)}$. The Lambert projection conserves the areas, $d\Omega=d\phi dc_\theta=d^2a=du dv$.]}
\end{figure}

Other reactions are expected to yield less events in 
water Cherenkov (as Kamiokande-II, IMB, Super-Kamiokande) 
or scintillators (as Baksan, LVD, KamLAND). This is true, e.g., for the elastic
scattering reaction, where the cross section 
$\sigma_{es}\sim G_F^2 m_e E$ is much smaller since $E\gg m_e$. 
However the electrons are scattered by 
 supernova neutrinos in the same direction of the 
incoming neutrinos, for the same reason, $E\gg m_e$. 
In other words, elastic scatterings will allow  
`neutrino imaging' as is illustrated in the figure, where we show the 
directions of arrival of the supernova events.
The cluster of events in the center, due to elastic scattering, is well visible over the background of
inverse beta decay events, which are only mildly directional. The statistical analysis shows that the direction of the supernova is determined with an accuracy of few degrees. See\cite{Pagliaroli:2009qy,ns,Ianni:2009bd} for further discussion.

We close this introductory note remarking that supernova neutrinos  
are a special target of neutrino astronomy. Indeed,
despite the rarity of the observable core collapses,  these neutrino 
can be certainly detected as demonstrated by SN1987A; moreover, they will permit to 
shed light on many open theoretical problems, regarding several fields: 
astrophysics, nuclear physics and particle physics.

\vskip5mm
\paragraph{Model of Neutrino Emission}
We describe now the model for the neutrino flux. 
Each emission phase is characterized by its intensity,
its duration and the average energy of the emitted neutrinos. So,
we have $6$ astrophysical parameters, namely: 
the initial mass ($M_a$), the time scale ($\tau_a$) and the
initial temperature ($T_a$) for the accretion
phase; the radius
($R_c$), the time scale ($\tau_c$) and the initial temperature
($T_c$) for the cooling phase. 
Moreover, in order to take into account the delay of the first observed event with respect to the
first neutrino, we introduced other free parameters, called {\it offset times}. Since 
the clocks of Kamiokande-II, IMB and Baksan were
not synchronized, we need 3 offset times.
For details, see\cite{Pagliaroli:2008ur}.

The expectations on these astrophysical parameters are: $R_c \sim
R_{ns} = 10-20$~km, $T_c=3-6$~MeV and the duration of the cooling
phase should be few (or many) seconds. The accretion phase has
$M_a < 0.6\ M_\odot$, $T_a$ is the few MeV range and the accretion
should last just $\sim 0.5$~s. The
accretion $\bar\nu_e$ luminosity can be estimated by:
\begin{equation}
L_{accr}\sim N_n \langle \sigma_{e^+ n} \rangle T_a^4
 \sim 5\times 10^{52} \label{accr}
\frac{\mbox{erg}}{\mbox{sec}}
\left( \frac{M_a}{0.1 M_\odot}\right)
\left( \frac{Y_n}{0.6} \right)
\left( \frac{T_a}{2\mbox{ MeV}} \right)^{\!6},
\end{equation}
where $N_n$ is the number of neutrons, that can be expressed in
term of an accretion mass, $M_a$, and of the fraction of neutrons
in the environment, $Y_n$. The cooling $\bar\nu_e$ luminosity has
instead the form:
\begin{equation}
L_{cool}\sim R_c^2 T_c^4 \sim 5\times 10^{51}
\frac{\mbox{erg}}{\mbox{sec}}
\left( \frac{R_c}{10\mbox{ km}} \right)^{\!2}
\left( \frac{T_c}{5\mbox{ MeV}} \right)^{\!4}, \label{cool}
\end{equation}
where $R_c$ and $T_c$, are, respectively radius and temperature of
the cooling. These formulae make evident that the two phases are
described by very different physical models, namely, by a
transparent atmosphere and by an opaque (black body) radiator;
this is the reason why the first phase is much more luminous.

The two phases are not contemporaneous. We parameterize the
$\bar\nu_e$ flux as follow:
\begin{equation}
\phi_{\bar \nu_e}(t) = \phi_a + (1-j_k(t))\times \phi_c (t-\tau_a).\label{parflux}
\end{equation}
Above $j_k$ represents a function that terminates the accretion
phase at $t \approx \tau_a$, approximated by
$\exp{[-(t/\tau_a)^2]}$. The flux is dominated by
the accretion phase at $t \ll \tau_a$, whereas the cooling phase
begins at $t \approx \tau_a$ and eventually dominates the flux
when $t\gg \tau_a$. With previous considerations in mind, we understand
that Eq.~(\ref{parflux}) describes the passage from an emission of volume to an emission 
of surface. The spectrum is quasi-thermal at any time; furthermore, the neutrinos luminosity and 
their average energies are smooth functions of the time.

The best fit values of the astrophysical parameters of accretion and cooling emission phases are found to be:
\begin{center}
\begin{tabular}{ccc}
$R_c=16^{+9}_{-5}\mbox{ km}$ &
$T_c=4.6^{+0.7}_{-0.6}\mbox{ MeV}$,  &
$\tau_c=4.7^{+1.7}_{-1.2}\mbox{ s}$, \\
$M_a=0.22^{+0.68}_{-0.15}\ M_\odot$, &
$T_a=2.4^{+0.6}_{-0.4}\mbox{ MeV}$,  &
$\tau_a=0.55^{+0.58}_{-0.17}\mbox{ s}$.
\end{tabular}
\end{center}
The  large errors are due to the limited statistics. The
results are close to what we expect from the
standard collapse, in particular we get ${\cal E}_{b}=2.2\times
10^{53}$ erg. There is a 2.5 $\sigma$ evidence for the accretion
phase. The 11 early events of Kamiokande-II (6), IMB (3) and Baksan (2) 
have a great probability to belong to the accretion phase.\footnote{
We add a remark on backgrounds events. The $6^{th}$-event of
Kamiokande-II, with  energy below $7.5$~MeV, has a probability of
$85\%$ to be due to background; a posteriori,
it should not be attributed to accretion.
Similarly, we found that the $13^{th}-16^{th}$ events
are almost surely due to background and there is still some chance of 
another background event. Similarly in Baksan, where 
the number of events is larger than expected and 
this is quite likely a priori.
In absence of more precise information, we assumed
that IMB was  background free; we checked that 
 the inferences do not change significantly assuming
that it had a background rate similar to the one of Kamiokande-II.}

\paragraph{Gravitational Waves and Neutrinos}

Gravity Waves (GW) are predicted by general relativity. They have
not been observed directly yet, but we will have soon detectors of enhanced
sensitivity. Core collapse supernovae can emit GWs
during the collapse (or during the explosion) of a core
collapse SN due to the change of the quadrupole moment of the star
structure. Recent simulations show that a gravitational signal is
emitted when the collapse of the inner core halts, as dictated by
the stiffening of the equation of state at nuclear density. The
consequent bounce of the outer core is pressure dominated without
strong influence of the rotation. Therefore, it is possible to
define a generic GW waveform which exhibits a positive pre-bounce
rise and a large negative peak, followed by a ring-down; so the
time of the bounce is strongly correlated to the time of the
maximum amplitude of the gravitational signal.

The duration of
the GW signal is about 10 ms. Therefore, to help the search of such
signals, one would like to identify the time of the bounce with an
error of the same order studying other types of signal emitted
from this event.
In ref.\cite{Pagliaroli:2009qy} it was argued that it is possible
to identify the time of the bounce within $\sim 10$ ms by an
analysis of the $\bar \nu_e$ signal from the explosion of a
galactic core collapse supernova; i.e., neutrinos can provide the required 
trigger for the search of GW.

In fact, extensive simulations of core
collapse SNe shows that the onset of $\bar \nu_e$ luminosity is
closely related to the time of the bounce. 
The time of the bounce $T_{\mbox{\tiny bounce}}$ can be determined by the following equation,
where the times in uppercase are absolute times, in UT, whereas those in
lowercase are relative intervals of time:
\begin{equation}
T_{\mbox{\tiny bounce}} = T_{\mbox{\tiny 1st}} - (t_{\mbox{\tiny
GW}} +t_{\mbox{\tiny mass}} \pm t_{\mbox{\tiny fly}}
+t_{\mbox{\tiny off}} ), \label{GW}
\end{equation}
 $T_{\mbox{\tiny 1st}}$ is the time of the first
neutrino event detected.
The time $t_{\mbox{\tiny GW}}$ is the interval between the
bounce of the outer core on the inner core and the beginning of
$\bar\nu_e$ emission. This is reliably known and ranges within
$t_{\mbox{\tiny GW}} = 1.5-4.5$~ms. The time $t_{\mbox{\tiny
mass}}$ is the delay, due to neutrino mass, between the arrival of
GW and neutrino signal; if we impose the
cosmological bound $\sum_i m_{\nu_i}<0.7$~eV, this is
negligible. The time interval $t_{\mbox{\tiny fly}}$ is the time
of fly between the two detectors and depends on the SN position in
the sky. Finally the non-negative parameter $t_{\mbox{\tiny off}}$
is the difference in time between the first neutrino and the first
event detected. In summary, the main terms in Eq.~(\ref{GW}) are
the fly time $t_{\mbox{\tiny fly}}$ and the offset (or response)
time $t_{\mbox{\tiny off}}$.

For the analysis of a future galactic supernova event, it is
important to consider the finite rising time of the $\bar\nu_e$
signal. This can be done multiplying the flux of Eq.~(\ref{parflux}) by the function $f_r
= 1-e^{-t/\tau_r}$. The rising
time $\tau_r \approx 50-150$~ms is a very important and new parameter of the
astrophysical model. It is related with the initial production of
$\bar \nu_e$ and depends strongly on the velocity of the shock
wave.

How to determine experimentally the time of fly from
neutrinos? If we know astronomically the direction of the SN, it
is easy to correct for the difference of arrival times. But even
if we do not know it, we can rely on elastic scattering (ES)
events, that are directional and suffice to determine the
time of fly precisely enough. (Note in particular that this 
can be applied to a supernova {\em without} optical output).

Thus, the problem reduces to the estimation of $t_{\mbox{\tiny
off}}$, i.e., the delay of the response of the detector to the
neutrino signal. If we have enough data and if we reconstruct at
the same time $\tau_r$, it is possible to reconstruct successfully
$t_{\mbox{\tiny off}}$ by fitting the data to the expectations. As
we can see from Table 3 of the reference\cite{Pagliaroli:2009qy},
these conditions are  expected to be satisfied for a galactic
supernova event: The response time and its error are correctly
estimated by the analysis. We conclude that the future galactic
supernova can provide us a precious information to test a key
prediction of general relativity and note incidentally, again
from\cite{Pagliaroli:2009qy}, that also the other astrophysical
parameters will be reconstructed with very good precision.

\paragraph{Neutrino Mass}
It is known since Zatsepin\cite{z} that supernova neutrinos permit
us to investigate the absolute mass scale of neutrinos. Several
works attempted this after SN1987A: see the Table for a partial
review\cite{cg,ll,cg1}. We will mostly discuss the last two entries of
this Table, but before doing that, we recall the idea.

\begin{table}[t]
\begin{center}
{\footnotesize
\begin{tabular}{ccc}\label{table1}
  Reference & $m_\nu$ & CL\\
  \hline
  Nat 326, 476 & $\le 11$ eV & -- \\
  Nat 329, 689 & O($13$) eV & -- \\
  PRL 58, 1906 & $\le 12$ eV & -- \\
  PRL 58, 2722 & $\le 26$ eV & -- \\
  PRL 59, 1864 & $\le 5$ eV & -- \\
  PLB 200, 366 & $\leq 16$ eV & 95\% \\
  PRD 35, 3598 & $\le 13.5$ eV 
  &   $1\sigma$ \\
  ELett 4, 953 & $\le 5.7$ eV & -- \\
  MPLA 2, 905  & $3.4 \pm 0.6$ eV & -- \\
  CNPP 17, 239 & $\le 30$ eV & -- \\
  BIHEP-CR-87-01 & $4.5 \pm 0.9$ eV & -- \\
  DTP/87/12 & $3.4 \pm 0.5$ eV & -- \\
  NPB 299, 734 & $m^2={4^{+28}_{-63}}$ eV$^2$ & --\\
  PLB 196, 259 & $\le 10$ eV & -- \\
  PRD 41, 682 & $\le 14$ eV & 95\% \\
  NPB 437, 243 & $\le 19.6$ eV & 95\% \\
  PRD 65, 063002 & $\le 5.7$ eV & 95\% \\
  arXiv:1002.3349 & $\le 5.8$ eV & 95\%
\end{tabular}}
\end{center}
\caption{Review of neutrino mass bounds in the
literature\label{table1}} \vskip-6mm
\end{table}

The flux, Eq.~(\ref{parflux}), is a function of the emission time $t_i$, namely
the time measured from the beginning of antineutrino emission. In terms of the 
absolute emission times we have $t_i=T_i^e-T_0^e$, that can be rewritten 
taking into account the velocity of the neutrino $v_i$ and the absolute detection time 
$T^d_i$ as follows:
\begin{equation}
t_i=\left(T^d_i -\frac{D}{v_i}\right)
-\left(T^d_0-\frac{D}{c}\right)\approx
\left(T^d_i - T^d_1\right) + \left(T^d_1- T_0^d\right) - 
 \frac{D}{2 c}  \left(\frac{m_\nu c^2}{E_{\nu,i}}\right)^2
\end{equation}
where $T_0^d$ is the minimum, possible detection time.
The first term in the r.h.s.\  $ T^d_i-T^d_1  \equiv  \delta
t_i$ is known from the experiment.
The second one,  $T_1^d-T_0^d \equiv t_{\mbox{\tiny off}}$ is by definition the
offset time, namely the delay between the first observed event and
the moment when the first neutrino has possibly reached the detector. 
The last one, denoted
by $\Delta t_i$, describes the effect of the finite neutrino mass.
Its numerical expression is:
\begin{equation}
\Delta t_i=2.6\mbox{ sec} \times \frac{D}{\mbox{50 kpc}}
\left(\frac{m_\nu c^2}{\mbox{10 eV}} \right)^2
 \left(\frac{\mbox{10 MeV }}{E_{\nu,i}} \right)^2. \label{pot}
\end{equation}
Putting together the above definitions, 
the emission time of each event, that enters the likelihood through the antineutrino flux, 
will be written as:
\begin{equation}
t_i=\delta t_i + t_{\mbox{\tiny off}}-\Delta t_i. \label{1}
\end{equation}
Again, the first term in
the r.h.s.\ is the relative time between
the $i$-th and the first observed event in the considered
detector, known directly from the data without significant error.
The last two terms, instead, have to be estimated by fitting the
data; both of them are positive (and thus lead to some
cancellation) but depend in a different way from the neutrino energy,
see Eq.~(\ref{pot}).
$E_{\nu, i}$, in turn,  can be inferred 
from the measured energy of the positron, $E_i$, 
which is known up to its
error $\delta E_i$.

From our statistical analysis\cite{cg1}, we obtain
from SN1987A data the bound
\begin{equation}
m_\nu<5.8\mbox{ eV at 95\% CL}.\label{bondi}
\end{equation}
The existence of a phase of accretion of about 0.5 s explains why
we are able to probe such neutrino masses. From
Eq.~(\ref{pot}) it should be evident that the most important
information to determine the mass is contained in the first, low
energy events. This consideration selects as most relevant the events of
Kamiokande-II, which incidentally, are also the events that probe
the existence of the accretion phase. The astrophysical
uncertainties are not very relevant; instead, the fact that offset
time is unknown contributes to worsen the bound. Note that the
agreement of the last two entries of the table\cite{ll,cg1} is to a large
extent accidental; e.g., the bound changes by $5-10$\% adopting
other conventional statistical procedures.

By mean of simulated
data, we demonstrated the 
possibility to probe sub-eV neutrino masses with 
Super-Kamiokande after a future galactic supernova. 
As discussed in\cite{cg1}, this limit requires a very precise knowledge of the
time and it is subject to strong fluctuations, related to the
position of the first low energy event.

\paragraph{Remark on the Relevant Time Scales}
We would like to summarize here the above results emphasizing the role of the relevant `time scales':\\
1) We recalled that the conventional astrophysical picture of the
explosion contains a relatively short time scale, namely the
duration of the accretion $\tau_a\sim 0.5$ s. We argued that
SN1987A data already provide some evidence for this time scale,
thanks to the fact that almost 40\% of the observed events fall in
this phase.\footnote{This is a significant extension of the usual
approach to SN1987A data analysis (as summarized, e.g., in Bahcall book) where the neutrinos are
thought to originate from a smooth, thermal emission with a much longer
time scale, $\tau_c=$ several seconds.}
A future galactic supernova will give us a much better determination of the
astrophysics and plausibly also of 10 ms (or even few ms) structures in the signal.\\
2) The search for GW can profit of the detection of neutrinos. We
remarked that the observation of a future galactic supernova
neutrino event could permit a determination with 10 ms precision
of the moment of the bounce, when a burst of GW is plausibly
emitted. (The same cannot be done for SN1987A, since the absolute
time of the events is unknown, except for IMB; thus, the bounce
should have  occurred 0.76 s before the first event seen by IMB,
see Eq.~(32) of \cite{Pagliaroli:2008ur}). The postulated
rise time of the signal, of the order of 50-150 ms, can also be measured.\\
3) The determination of the neutrino mass has to go through a
precise determination of times: see Eq.~(\ref{1}); the smaller the time scale we probe, 
the better the limit on the mass we obtain. E.g., the bound from
SN1987A (several eV) is essentially determined by the existence of an accretion
phase, while for a future supernova the relevant time scale will be the 
rise time of the signal (probing the sub-eV region).
This could be further improved detecting a hypothetical shorter burst 
of $\bar\nu_e$, lasting only few ms. This is not
expected to exist for a standard collapse, but it has been found
in the first numerical simulations concerning the formation of a
hybrid star\cite{sss}.

\vskip5mm
\paragraph{Conclusions}
The problem to explain SN explosion is still open, however, a
reference (standard) model does exist. In this work, we focussed
on this standard scenario and discussed possible observational
tests.

The SN1987A data present unexpected features, however,
KII, IMB and Baksan data fit in a specific model. They show a hint
of an initial high-luminosity phase, as the one expected for a 
standard neutrino emission. A future galactic SN will
permit much more precise tests providing a huge amount of
new information; remarkably, it will also give the GW burst timing
within $\sim 10$ msec. Also it will be possible to establish a
relatively tight neutrino mass bound, up to the sub-eV range.

Finally, we noted that many of the interesting results on astrophysics
and particle physics 
that can be obtained by 
observations of supernova neutrinos are essentially based on
precise measurements of time. This consideration emphasizes the
crucial importance to improve our knowledge on the astrophysics of the
neutrino emission.

\paragraph*{Acknowledgments}
{\small FV thanks R.~Ruffini for the invitation and 
ICRANet, Pescara for hospitality during the preparation of this work;
FRT is grateful to CAPES and FAPESP for the financial support;
GP acknowledges the grant of POR Abruzzo (FSE).}


\begin{thebibliography}{0}


\bibitem{nad}
D.K. Nadyozhin,
Astrophysics and Space Science {\bf 53}, 131 (1978).

\bibitem{Bethe:1984ux}
  H.~A.~Bethe and J.~R.~Wilson,
  Astrophys.\ J.\  {\bf 295}, 14 (1985).

\bibitem{Pagliaroli:2008ur}
  G.~Pagliaroli, F.~Vissani, M.~L.~Costantini and A.~Ianni,
  Astrop.\ Phys.\  {\bf 31}, 163 (2009).



\bibitem{ll}
  T.J.~Loredo and D.Q.~Lamb,
  Phys.\ Rev.\  D {\bf 65} (2002) 063002.


\bibitem{Strumia:2003zx}
  A.~Strumia and F.~Vissani,
  Phys.\ Lett.\  B {\bf 564}, 42 (2003)
  [arXiv:astro-ph/0302055].



\bibitem{Ianni:2009bd}
  A.~Ianni {\em et al.} 
  Phys.\ Rev.\  D {\bf 80}, 043007 (2009).


\bibitem{jcap}
M.~L.~Costantini, A.~Ianni, G.~Pagliaroli and F.~Vissani,
  JCAP {\bf 0705}, 014 (2007).

\bibitem{k}
K.S.~Hirata {\it et al.},
  Phys.\ Rev.\  D {\bf 38} (1988) 448;
K.~Hirata {\it et al.}  
  Phys.\ Rev.\ Lett.\  {\bf 58} (1987) 1490.

\bibitem{i}
  R.M.~Bionta {\it et al.},
  Phys.\ Rev.\ Lett.\  {\bf 58} (1987) 1494;
C.B.~Bratton {\it et al.}  
  Phys.\ Rev.\  D {\bf 37} (1988) 3361.

\bibitem{b}
E.N.~Alekseev, L.N.~Alekseeva, V.I.~Volchenko and I.V.~Krivosheina,
  JETP Lett.\  {\bf 45} (1987) 589 and 
  Phys.\ Lett.\  B {\bf 205} (1988) 209.


\bibitem{Pagliaroli:2009qy}
  G.~Pagliaroli, F.~Vissani, E.~Coccia, W.~Fulgione,
  Phys.\ Rev.\ Lett.\  {\bf 103}, 031102 (2009).

\bibitem{ns}
G. Pagliaroli, F.L. Villante, F. Vissani, 
Nuovo Saggiatore 25, no. 3-4 
(2009) 5-19. 

\bibitem{onlus}
  F.~Vissani, G.~Pagliaroli and F.~L.~Villante,
Il Nuovo Cimento C 32 (2009) 353.

\bibitem{z}
  G.T.~Zatsepin,
  Pisma Zh.\ Eksp.\ Teor.\ Fiz.\  {\bf 8} (1968) 333.

\bibitem{cg}
  J.~N.~Bahcall and S.~L.~Glashow,
  Nature {\bf 326}, 476 (1987).
  L.~M.~Krauss,
  Nature {\bf 329}, 689 (1987).
  W.~D.~Arnett and J.~L.~Rosner,
  Phys.\ Rev.\ Lett.\  {\bf 58}, 1906 (1987).
  K.~Sato and H.~Suzuki,
  Phys.\ Rev.\ Lett.\  {\bf 58}, 2722 (1987).
  J.~Arafune, M.~Fukugita, T.~Yanagida and M.~Yoshimura,
  Phys.\ Rev.\ Lett.\  {\bf 59}, 1864 (1987).
  S.~H.~Kahana, J.~Cooperstein and E.~Baron,
  Phys.\ Lett.\  B {\bf 196}, 259 (1987).
  E.~W.~Kolb, A.~J.~Stebbins and M.~S.~Turner,
  Phys.\ Rev.\  D {\bf 35}, 3598 (1987).
  M.~Roos,
  Europhys.\ Lett.\  {\bf 4}, 953 (1987).
  H.~Huzita,
  Mod.\ Phys.\ Lett.\  A {\bf 2}, 905 (1987).
  D.~N.~Schramm,
  Comments Nucl.\ Part.\ Phys.\  {\bf 17}, 239 (1987).
  C.~M.~Xu, X.~J.~Wu and T.~P.~Li, BIHEP-CR-87-01 (1987).
  D.~Evans, R.~Fong and P.~D.~B.~Collins, DTP/87/12 (1987).
  L.~F.~Abbott, A.~De Rujula and T.~P.~Walker,
  Nucl.\ Phys.\  B {\bf 299}, 734 (1988).
  D.~N.~Spergel and J.~N.~Bahcall,
  Phys.\ Lett.\  B {\bf 200}, 366 (1988).
  F.~T.~Avignone and J.~I.~Collar,
  Phys.\ Rev.\  D {\bf 41}, 682 (1990).
  P.~J.~Kernan and L.~M.~Krauss,
  Nucl.\ Phys.\  B {\bf 437}, 243 (1995).
\bibitem{cg1}
G.~Pagliaroli, F.~Rossi Torres, F.~Vissani,
arXiv:1002.3349, Astrop. Phys. (2010).

\bibitem{sss} 
I.~Sagert {\it et al.},
  Phys.\ Rev.\ Lett.\  {\bf 102}, 081101 (2009).

\end{thebibliography}
\end{document}